# Apache Lucene as Content-Based-Filtering Recommender System: 3 Lessons Learned

Stefan Langer[1] and Joeran Beel[2,3]

[1]Otto-von-Guericke University, Department of Computer Science, Magdeburg, Germany
langer@ovgu.de
[2]Trinity College Dublin, Department of Computer Science, ADAPT Centre, Ireland
joeran.beel@adaptcentre.ie
[3]National Institute of Informatics, Digital Content and Media Sciences Research Division, Tokyo, Japan
beel@nii.ac.jp

**Abstract.** For the past few years, we used Apache Lucene as recommendation framework in our scholarly-literature recommender system of the reference-management software Docear. In this paper, we share three lessons learned from our work with Lucene. First, recommendations with relevance scores below 0.025 tend to have significantly lower click-through rates than recommendations with relevance scores above 0.025. Second, by picking ten recommendations randomly from Lucene's top50 search results, click-through rate decreased by 15%, compared to recommending the top10 results. Third, the number of returned search results tend to predict how high click-through rates will be: when Lucene returns less than 1,000 search results, click-through rates tend to be around half as high as if 1,000+ results are returned.

## 1 Introduction

Apache Lucene/Solr is probably the most common search framework, and it is frequently used by content-based-filtering recommender systems (Bancu et al., 2012; Caragea et al., 2014; Garcia Esparza, O'Mahony, & Smyth, 2010; Jonnalagedda, Gauch, Labille, & Alfarhood, 2016; Livne, Gokuladas, Teevan, Dumais, & Adar, 2014; Mitzig et al., 2016; Phelan, McCarthy, & Smyth, 2009; Pohl, 2007; Pursel et al., 2016; Shelton, Duffin, Wang, & Ball, 2010). Lucene's build-in recommendation method, which uses a classic TF-IDF-weighted term-vector retrieval approach, is also generally used frequently as a baseline method that typically achieves good results. (Demner-Fushman et al., 2011; Gipp, Meuschke, & Lipinski, 2015; Schwarzer et al., 2016)

We used Lucene to implement a research-paper recommender system in *Docear* (Beel, Gipp, Langer, & Genzmehr, 2011; Beel, Gipp, & Mueller, 2009; Beel, Langer, Gipp, & Nürnberger, 2014; Beel, Langer, Genzmehr, & Müller, 2013; Beel, Langer, Genzmehr, & Nürnberger, 2013). Docear is a free and open-source reference manager, comparable to tools like Endote, Zotero, Mendeley, or Citavi. Docear has approximately 50,000 registered users and uses mind-maps to manage PDFs and references. Since 2012, Docear has been offering a recommender system for 1.8 million publicly available research papers on the web. Recommendations are displayed as a list of ten research papers, showing the title of the recommended papers (**Fig. 1**). Clicking a recommendation opens the paper's full-text (PDF) in the user's web browser. Between 2012 and 2015, the recommender system delivered around one million recommendations to more than 7,000 researchers. For more details on the recommender system please refer to Beel et al. (2014).

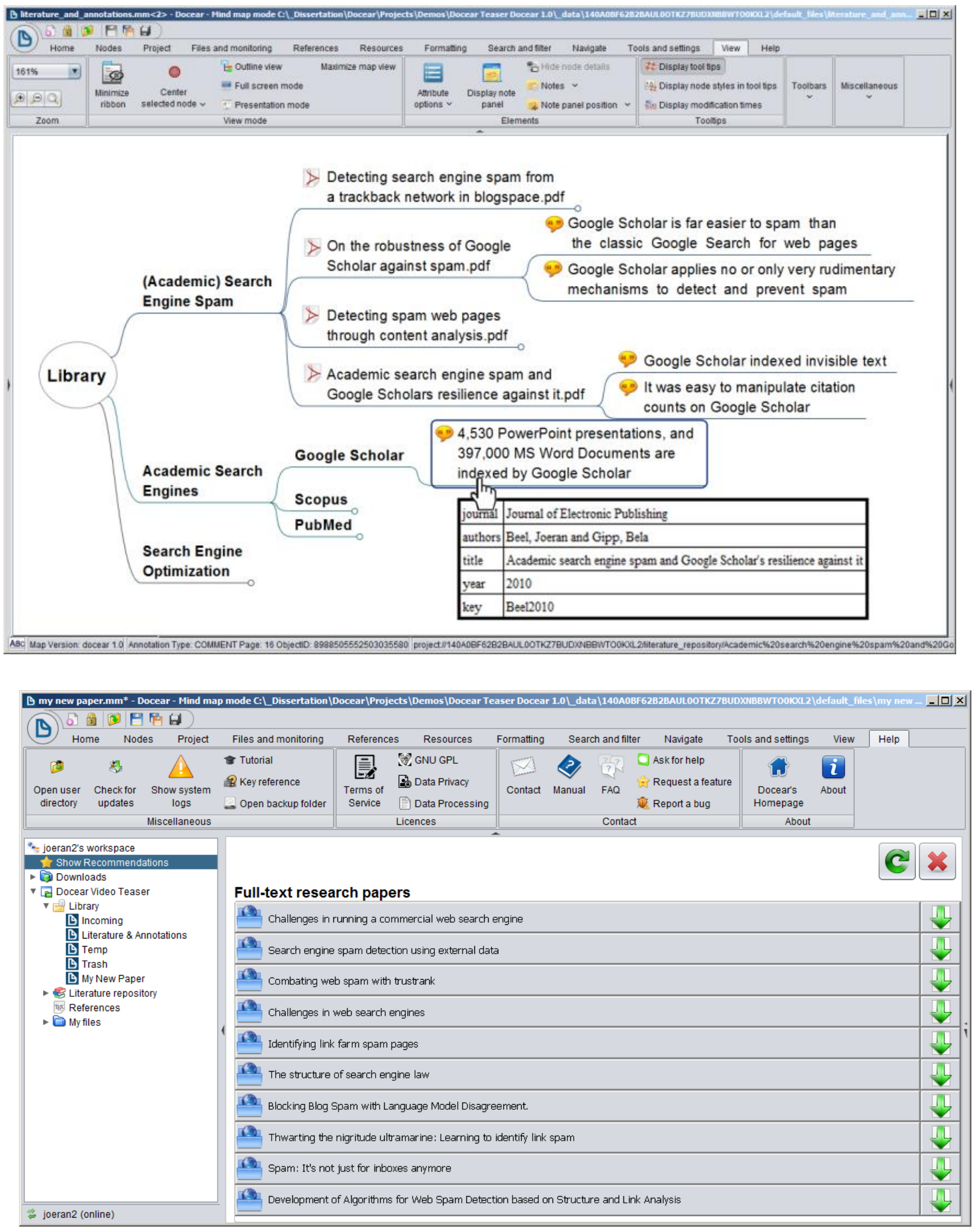

**Fig. 1.** Screenshots of Docear and the recommender system

In this paper, we share some experiences we made with Lucene, focusing on three aspects. First, we analyze the meaning of Lucene's relevance scores. Second, we analyze how effective recommendations are based on Lucene's suggested rank. Finally, we analyze the relationship between the amount of recommendation candidates that Lucene returns and the recommendation effectiveness. Although we did our research in the context of research-paper recommendations, results might also be interesting for other recommender-systems domains that use Lucene, for instance, in the domains of news recommender systems, website recommender systems, or tweet recommender systems (Chen, Ororbia, Alexander, & Giles, 2015; Duma, Liakata, Clare, Ravenscroft, & Klein, 2016; Garcia Esparza et al., 2010; Jonnalagedda et al., 2016; Mitzig et al., 2016; Phelan et al., 2009; Shelton et al., 2010).

## 2 Methodology

All presented results are based on data that we collected between May 2013, and October 2014. During this time, Docear's recommender system delivered 418,308 recommendations to 4,674 unique users. We use click-through rate as measure for the effectiveness of delivered recommendations. Click-through rate (CTR) describes the ratio of clicked and delivered recommendations. For more details on click-through rate and its suitability as evaluation metric please refer to Beel & Langer (2015). All reported differences are statistically significant ($p < 0.05$) based on a two-tailed t-test.

## 3 Results & Discussion

### 3.1 Lucene's Relevance Scores

Lucene provides relevance scores for each recommendation. This information could be used, theoretically, to recommend only documents with a relevance score above a certain threshold. However, on the Web it is often reported that these scores cannot be used to compare relevancies of recommendations between different queries, or to conclude from the relevance score how relevant the search result or recommendation is overall.[1] Our data shows a slightly different picture.

In our data, the highest relevance score for a recommendation was 19.01, median was 0.16 and mean was 0.22. **Fig. 2** shows that CTR was lowest (3.36%) for recommendations with a relevance score below 0.01, and highest (6.16%) for relevance scores of 1 and above. For recommendations with relevance scores between 0.1 and 0.8, CTR remained mostly stable around 5%. Overall, there is a notable trend: CTR increases, the higher Lucene relevance scores become.

Our observation contradicts the common claims that Lucene's relevance score cannot be used to estimate a search result's absolute relevance. If, for instance, an operator of a recommender system decided that a click-through rate of at least 4% was desirable, then recommendations with a relevance score below 0.25 should probably be discarded.

[1] https://wiki.apache.org/lucene-java/ScoresAsPercentages

Similarly, our result might lead to the conclusion to only recommend documents above a certain relevance threshold, e.g. 1. However, recommending only documents with a relevance score of 1 and above is probably not sensible as only a small fraction of recommendations had a relevance score of 1 and above (0.60%). Similarly, it might seem sensible to not recommend documents with relevance scores below 0.025 as these documents had very low CTRs. However, only a small fraction of recommendations (4.27%) had relevance scores below 0.025, so this decision would barely affect the overall click-through rate.

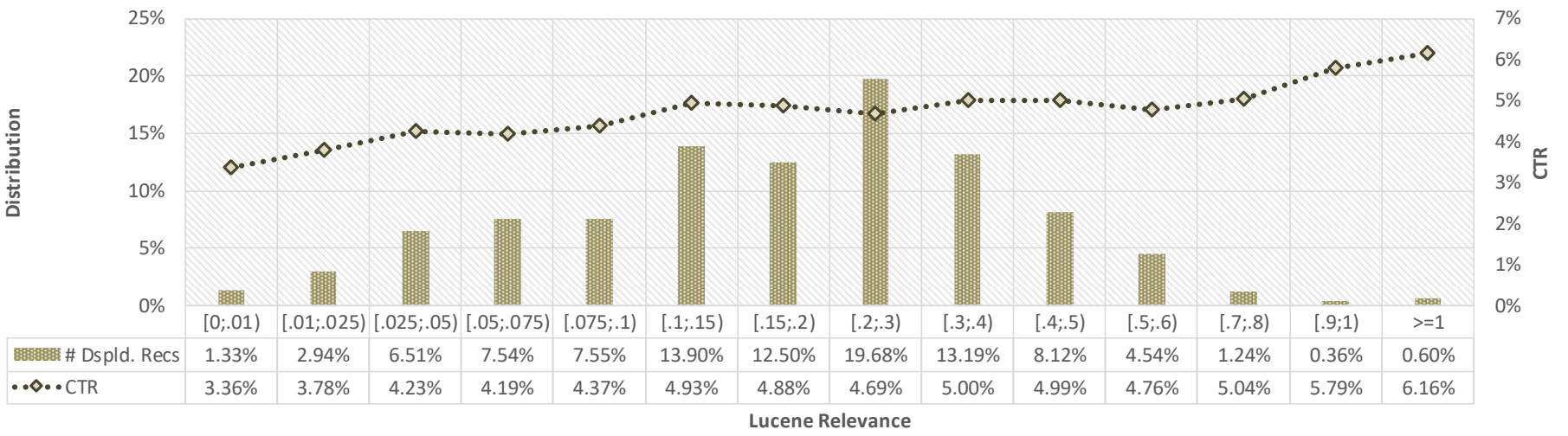

**Fig. 2.** Lucene relevance score and corresponding CTR

## 3.2 Lucene's Rank

To increase the diversity of recommendations, Docear's recommender system randomly chose 10 recommendations out of the top50 results returned by Lucene. However, this leads to lower click-through rates. Recommendations originally being ranked 1 by Lucene received CTRs of 6.83% on average and recommendations on rank 2 received CTRs of 6.08% on average (**Fig. 3**). For ranks 3 to 10, CTR remains stable around 5.3% and then CTR constantly decreases the lower the original rank.

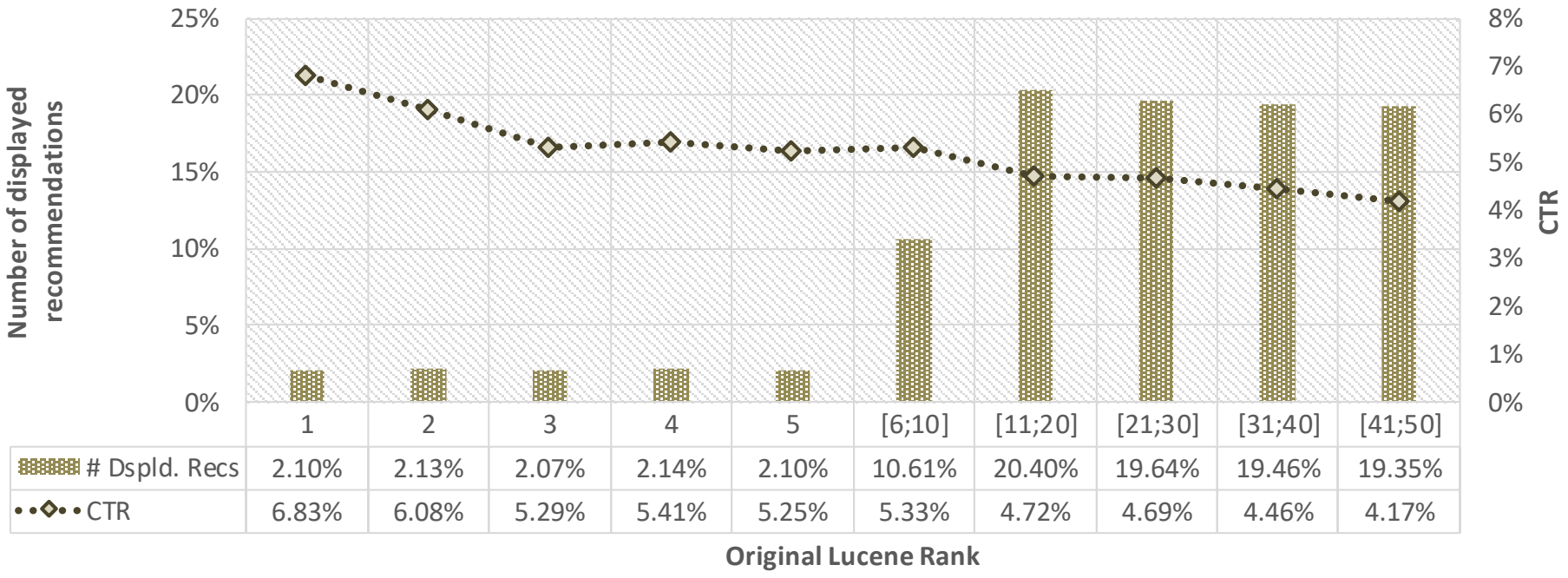

**Fig. 3.** Lucene's rank and corresponding CTR

Overall, recommendations being in Lucene's top10 results, achieved CTRs of 5.55% on average, while the top50 achieved CTRs of 4.73% on average. This means, selecting randomly 10 recommendations from the top50 candidates decreases recommendation effectiveness by around 15%, compared to showing recommendations from the top10 only. The recommender system shuffled recommendations before they were displayed. This means, position bias cannot have influenced the results (Craswell, Zoeter, Taylor, & Ramsey, 2008; Hofmann, Schuth, Bellogin, & Rijke, 2014; Pan et al., 2007; Wang, Bendersky, Metzler, & Najork, 2016).

### 3.3 Number of Recommendation Candidates

By default, Lucene returns 1,000 recommendations, i.e. search results. In our data, Lucene returned the maximum possible amount of 1,000 results for 91.25% of all term-based recommendations (**Fig. 4**). In contrast, only for 0.05% of citation-based searches 1,000 results were returned. Most citation-based searches returned between one and nine results (34.84%) or between 10 and 24 results (29.94%). Click-through rates seem to be rather high when only few results were returned. For term-based searches, results are the opposite: the more recommendation candidates are available, the higher the CTR tends to be. Consequently, for term-based recommendations, the number of results might be a good approximation of recommendation effectiveness. If less than 1,000 results are returned it might make sense to no recommend the documents or try an alternative recommendation approach.

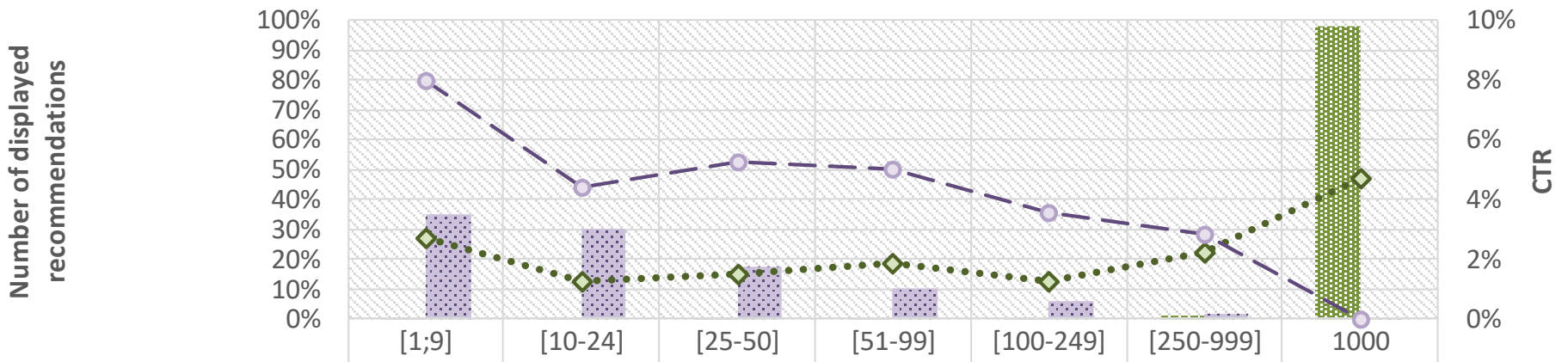

| | [1;9] | [10-24] | [25-50] | [51-99] | [100-249] | [250-999] | 1000 |
|---|---|---|---|---|---|---|---|
| # Dspld. Recs (Terms) | 0.18% | 0.21% | 0.18% | 0.19% | 0.38% | 0.92% | 97.94% |
| # Dspld. Recs (Cit.) | 34.84% | 29.94% | 17.37% | 10.28% | 5.93% | 1.59% | 0.05% |
| CTR (Terms) | 2.73% | 1.28% | 1.50% | 1.86% | 1.29% | 2.22% | 4.70% |
| CTR (Citations) | 7.96% | 4.43% | 5.27% | 5.05% | 3.54% | 2.84% | 0.00% |

Number of Recommendation Candidates in Lucene

**Fig. 4.** CTR based on the number of recommendation candidates in Lucene

## 4 Summary & Future Work

From our analysis, we learned three lessons. First, Lucene's relevance score allows to predict – to some extend – how relevant a recommendation will be for a user. For instance, in our scenario, it seems sensible to not recommend documents with a relevance score below 0.025. However, since only few recommendations had such a low rele-vance score, discarding them will probably not notably affect the overall recommendation effectiveness. Second, recommending ten recommendations out of the top50 results might be sensible. Although this process decreases the overall recommendation effectiveness by 15%, the recommendation diversity or number of total recommendations is increased. Third, the number of recommendation candidates returned by Lucene is suitable to approximate the recommendation effectiveness. If Lucene returns less than 1,000 results for term-based recommendations, the click-through rate probably will be around half as high as if 1,000 candidates are returned. In the case of less than 1,000 results, it might make sense to not display the recommendations or generate recommendations again with another recommendation approach.

For the future, we suggest to repeat our analyses in different scenarios, for instance, with news recommenders or other literature recommender systems, to see if Lucene behaves in the same way as in the scenario of Docear. Currently, we are developing a recommender system as-a-service named Mr. DLib (Beel & Gipp, 2017; Beel, Gipp, Langer, Genzmehr, et al., 2011). Mr. DLib will allow us to conduct such analyses with different partners.

## 5 Acknowledgements

This work was supported by a fellowship within the FITweltweit programme of the German Academic Exchange Service (DAAD). In addition, this publication has emanated from research conducted with the financial support of Science Foundation Ireland (SFI) under Grant Number 13/RC/2106.